\documentclass{revtex4}

\newcommand{\beq}{\begin{equation}}
\newcommand{\eeq}{\end{equation}}
\newcommand{\bea}{\begin{eqnarray}}
\newcommand{\eea}{\end{eqnarray}}
\newcommand{\ov}{\overline}

\begin{document}

\title{See-saw and Grand Unification}

\author{Goran Senjanovi\'c}
\address{ International Centre for Theoretical Physics, 34100 Trieste, Italy }

\begin{abstract}
I review the profound connection between the see-saw mechanism for neutrino
masses and grand unification. This connection points naturally towards SO(10)
grand unified theory. The emphasis here is on the supersymmetric theory, but I
also discuss salient features of its split supersymmetry version and ordinary 
non-supersymmetric SO(10). Particular attention is
paid to the crucial issue of the minimal such theory, i.e. the question of
the Higgs sector needed to break SO(10) down to the Minimal Supersymmetric
Standard Model or the Standard Model.  Some essential features of the see-saw
mechanism are clarified, in particular its precise origin at the high scale.
\end{abstract}

 \maketitle
\section*{Prelude}
 
   I have been asked by the organizers of the SEESAW25 to review
the see-saw mechanism in  connection with grand unification. 
Due to the enormous body of work in the field, 
neither time nor space allow me to do a complete job. Instead I focus
here on the work regarding pure grand unification, the work
connected with the search for the minimal grand unified theory, 
both supersymmetric and not. 

I am forced thus to omit  some important issues such as doublet triplet
splitting, grand unification in extra dimension, non minimal models, fermion
mass textures, and more. 
For some complementary reviews of these topics (and
not only) see:\cite{review,Altarelli:2004za,Albright:2002np,Pati:2004fu}.

\section{The parameters}

We know today \cite{cosmo} that neutrinos are massive and we know the two mass
differences that correspond to the atmospheric and solar neutrino oscillations
\beq
\Delta m_{A}^{2} \simeq (2.5 \pm 0.6)10^{-3} \mbox{eV}^{2} \, ; \quad \Delta
m_{\odot}^{2}
\simeq (8.2\pm 0.6)10^{-5} \mbox{eV}^{2}
\label{cero}
\eeq
 The corresponding mixing angles are
 \beq
 \theta_{A}= 45^\circ \pm 6^\circ  \, ; \quad \theta_{\odot} = 32.5^\circ\pm 2.5^\circ
\label{cero1}
 \eeq
 From (\ref{cero}), the mass of the heaviest neutrino has a lower limit
 \beq
 m_{\nu}^{\rm max} \geq 5 \times 10^{-2} \mbox{eV} \label{cero2}
 \eeq
 We also know from $\beta$ decay that 
 \beq
 m_{\nu_{e}} \leq   2.2 \mbox{eV} \; \; (95\% c.l.)
 \label{cero3}
 \eeq
 and from cosmological  data we know  that the sum of neutrino masses
is small
  \beq
 \sum m_{\nu} \lesssim   0.4 \mbox{eV}  - 1.7 \mbox{eV} \label{cero4}
 \eeq
  Thus, even if degenerate, neutrinos are very light, $m_{\nu} \lesssim (0.15
\mbox{eV} - 0.6 \mbox{eV})$.  How to understand so small neutrino masses? A simple
answer lies in the see-saw mechanism.

\section{See-saw in the Standard Model}

The see-saw mechanism\cite{seesaw} of neutrino
masses in the context of the Standard Model
(SM), is obtained by adding a right-handed neutrino and writing the most general
$SU(2)_{L}\times U(1)_{Y}$ Yukawa Lagrangean
\beq
 L_{y}(\nu) = y_{D} \ov{\nu}_{L} \, \phi \, \nu_{R} + \frac{1}{2}M_{R}
\,\nu_{R}^{T}\, C^{-1} \, \nu_{R} + h.c.
\label{uno}
\eeq 
In our symbolic notation, $y_{D}$ is a Yukawa matrix in flavor space. From
(\ref{cero}), at least two light neutrinos are massive, and thus one needs at
least two right-handed neutrinos. Having in mind grand unification with
quark-lepton symmetry {\it a la} Pati-Salam as  a natural framework to study
neutrino masses, in what follows I assume three right-handed neutrinos and
suppress generation indices. 

Now, (\ref{uno}) creates a mess, unless $M_{R} \gg y_{D} \langle \phi\rangle
$. This is quite natural though, since $M_{R}$ is a gauge invariant quantity
and thus expected to be very large: $M_{R} \gg M_{W}$.  The principle that
gauge invariant  quantities lie much above the scale of the breaking of the
symmetry in question lies at the heart of the see-saw mechanism. In what
follows, I will  stick to it consistently.  
 
  Since $y_{D} \leq 1$, with $M_{R} \gg
\langle \phi\rangle $ one gets automatically small neutrino masses
\beq
M_{\nu}^{I} = m_{D}^{T}\,
 M_{R}^{-1} \, m_{D} \label{dos}
 \eeq
where $m_{D} \equiv y_{D} \langle \phi\rangle $ and $I$ stands for the type I
see-saw, which has become the common name for this realization of small
neutrino masses. 

Alternatively, you could add to the SM a $SU(2)_{L}$ triplet $\Delta $, with
$B-L (\Delta) =2$, and the Yukawa couplings
\beq
L_{y}(\nu) = y_{\nu} \ell_{L}^{T} C \Delta \ell_{L} + h.c. \label{tres}
\eeq
where $\ell_{L}$ stands for the leptonic doublets.
The triplet gets a non-vanishing vev $
\langle \Delta \rangle \simeq   \frac{M_{W}^{2}}{M_{\Delta}} $
if the triplet mass $M_{\Delta} \gg M_{W}$. 
The same
principle as before ensures small neutrino masses
\cite{Lazarides:1980nt}
\beq
M_{\nu}^{II} \simeq y_{\nu} \frac{M_{W}^{2}}{M_{\Delta}} \label{seis}
\eeq
where the superscript $II$ stands for the type II see-saw as is commonly
called. 

In the SM  we cannot distinguish the
two mechanisms for $M_{R}, M_{\Delta} \gg M_{W}$. They are both simply the
expression of the effective operator analysis which tells us that the leading
$SU(2)_{L} \times U(1)_{Y}$ Yukawa coupling is of
dimension 5 \cite{Weinberg:1979sa}
\beq
L_{y}^{\rm eff}(\nu) = f \frac{1}{M}  \left(\ell^{T} \sigma_{2}\Phi
\right)  C \left( \phi^{T} \sigma_{2} \ell \right)
\label{siete}\eeq
where $M \gg M_{W}$ and $f$ is a matrix in generation space.  Hence
small, see-saw like suppressed neutrino masses
\beq
M_{\nu} \simeq f \frac{M_{W}^{2}}{M} \label{ocho}
\eeq
  Obviously,
both type I and type II see-saw are of the form (\ref{ocho}), as they have to
be.  There is no sense in trying to distinguish type I form type II in the SM,
not without any new physics being invoked. After all, if $y_{D} \propto
y_{\nu} \propto M_{R}$, we will even have the same flavor structure for
neutrino mass matrices.  In order to study this issue, we must go beyond the
SM. 

Our task is highly nontrivial. In order that the see-saw mechanism be tested,
in order that it be a theory, we need first of all to know the origin of the
mechanism: is it type I, type II or something else? \footnote{ It is
interesting that the only other alternative is the fermionic triplet (actually
triplets, at least two are needed similar to right-handed
neutrinos)\cite{Hambye:2003rt}.}   
Since $M_{R}$ has to be very large (unless Yukawa couplings are extremely
small), it is natural to consider grand unification as 
a framework of new
large mass scales and of  quark-lepton unification which sheds
light on $y_{D}$ (and/or $y_{\nu}$).

\section{How to incorporate see-saw in GUTs ?}

This question is intimately tied up with the fundamental issue of the choice
of {\it the} grand unified theory. After three decades of grand unification,
there is no consensus today of what the theory is.  

First, the symmetry reasoning. In the SM,
$B-L$ is an accidental, anomaly free $U(1)$ global symmetry. It is of
course broken by $M_{R}$  (or $M$ in the effective operator
language). If you believe in an accidental global $B-L$, then the SU(5)
direction is the natural one, since this phenomenon persists. 

On the other hand, the fact that  $B-L$ is anomaly free makes a strong case
for its gauging. This would in turn induce $(B-L)^{3}$ anomaly; the simplest
remedy is to introduce the right-handed neutrinos (one per generation). The
natural setting is then provided by Left-Right ($L-R$) symmetric theories
\cite{leftright} 
where right-handed neutrinos
are a must and $B-L$ has a simple physical interpretation from the electric
charge formula
\beq
Q = T_{3L} + T_{3R} + \frac{B-L}{2}
\label{qem}
\eeq
 The new scale $M_{R}$ is then the scale of
$SU(2)_{R}$, or better to say $L-R$ (parity) symmetry breaking.  It is
important to stress that both type I and type II see-saw emerge naturally in
this case and are deeply connected. This route points towards Pati-Salam
quark-lepton unification  and SO(10) as a grand
unified theory (GUT). 

In the next two sections, I go through both SU(5) and SO(10) supersymmetric
theories.   Of course, SU(5) needs low energy
supersymmetry  (or split supersymmetry\cite{split}), whereas the
same cannot be said of the
SO(10) theory. I will thus discuss both supersymmetric and ordinary SO(10).

\section{Grandunification: SU(5)}

As remarked before, SU(5) is a natural theory if you give up gauging $B-L$.
The minimal SU(5) theory fails, for the gauge couplings do not unify. With low
energy supersymmetry the couplings unify as predicted more than two decades
ago\cite{susyunif}.  

The minimal theory with only $5_{H}$ (and $\ov{5}_{H}$ in SUSY),
predicts $m_{d} = m_{e}$ at $M_{GUT}$, generation by generation
\cite{Chanowitz:1977ye}.
Whereas
$m_{b} = m_{\tau}$ works well, for other generations $ | m_{\mu}| \simeq 3
|m_{s}|$,  $ | m_{e}| \simeq 1/3
|m_{d}|$ is needed (again, at $M_{GUT}$). This is easily achieved without any
change in the theory, by adding higher dimensional operators suppressed by
$1/M_{Pl}$. The theory loses then its predictivity in determining precisely
$M_{GUT}$
and $\tau_{p}$, but is saved from being ruled out by a too fast $d=5$ proton
decay\cite{Bajc:2002bv}. Alternatively, one could add  more Higgs superfields,
say $45_{H}$ as in the Georgi-Jarlskog\cite{Georgi:1979df} approach. 

What about neutrino masses in $SU(5)$? You could choose from three simple
possibilities, none of them very appealing:
\begin{enumerate}
\item Add right-handed neutrinos, SU(5) singlets. In this case $M_{R}$ is a
gauge invariant quantity, and by the principle of naturalness $M_{R} \gg
M_{GUT}$. This is no good, since then $m_{
\nu} \ll M_{W}^{2}/M_{GUT} \simeq 10^{-3} \mbox{eV}$, which is too small to explain
the solar and especially the atmospheric neutrino data.

\item Add a  $SU(2)_{L}$ triplet as before, with $\Delta $
 contained in $15_{H}$, a two index symmetric Higgs superfield. With the same
principle as above, you reach the same conclusion.

\item You could  write a higher-dimensional operator {\it a la} Weinberg
\cite{Weinberg:1979sa}
\beq
O_{5} = f \frac{1}{M_{\nu}} \ov{5}_{F} \ov{5}_{F} {5}_{H} {5}_{H}
\eeq
with $M_{\nu} \gg M_{GUT}$, say $M_{\nu} \sim M_{Pl}$
\cite{Barbieri:1979hc}. 
Strictly speaking, (1) and (2) correspond to this, as in the SM case discussed
before. Again, the Planck scale suppression is too large to account for 
the atmospheric or solar neutrino data. Such terms can be relevant though 
for small splittings in the case of degenerate neutrinos\cite{Akhmedov:1992et}.
\end{enumerate}

All this does not prove that we must give up on SU(5). After all, we can
fine-tune $M_{R}$, the trouble is that
 the Yukawas are  arbitrary, as much as in the SM. Such a
theory, with all the tree-level and $1/M_{Pl}$ corrections to Yukawa
couplings needed to correct fermionic mass relations, has too many parameters.
In order to pursue this direction, one should go beyond SU(5) and invoke extra
family horizontal symmetries, discrete, global or local (for a review and
references see\cite{Altarelli:2004za}).
Instead, we turn to SO(10) which is tailor fit for a theory of fermion masses. 

\section{SO(10): the minimal theory of matter and gauge coupling unification}

There are a number of features that make SO(10) special:
\begin{enumerate}
\item a family of fermions is unified in a 16-dimensional spinorial
representation; this in turn predicts the existence of right-handed neutrinos
\item $L-R$ symmetry is a finite gauge transformation in the form of charge
conjugation. This is a consequence of both left-handed fermions $f_{L}$ and
its charged conjugated counterparts $(f^{c})_{L} \equiv C \ov{f}_{R}^{T}$
residing in the same  representation $16_{F}$. 
\item in the supersymmetric version,  matter parity $M= (-1)^{3(B-L)}$,
equivalent to the R-parity $R= M (-1)^{2 S}$,  is a gauge transformation
\cite{rparity},
a part of the center $Z_{4}$ of SO(10). It simply reads $16 \to -16$,
$10 \to 10$.  Its fate depends then on the pattern of symmetry breaking
(or the choice of Higgs fields);  it turns
out that in the  renormalizable version of the theory  R-parity 
 remains exact
at all energies\cite{lrsusy,Aulakh:2000sn}. The lightest supersymmetric
partner (LSP) is then stable and is a natural candidate for the dark matter of
the universe.
\item its other maximal subgroup, besides $SU(5) \times U(1)$, is $SO(4)
\times SO(6) = SU(2)_{L} \times SU(2)_{R} \times SU(4)_{c}$ symmetry of Pati
and Salam. It explains immediately the somewhat mysterious relations  $m_{d} =
m_{e}$ (or $m_{d} = 1/3 m_{e}$) of SU(5).   
\item the unification of gauge couplings can be achieved with or without
supersymmetry. 

\item the minimal renormalizable version (with no higher dimensional
$1/M_{Pl}$ terms) offers a simple and deep connection between $b-\tau$
unification and a large atmospheric mixing angle in the context of the type II
see-saw \cite{Bajc:2002iw}.
 
\end{enumerate}
In order to understand some of these results, and in order to address the
issue of construction of the theory, we turn now to the Yukawa sector. 

\subsection{Yukawa sector}

Fermions belong to the spinor representation $16_{F}$ 
\cite{Mohapatra:1979nn}. From
\beq
16 \times 16 = 10 + 120 + \ov{126}
\label{nueve}
\eeq
the most general Yukawa sector in general contains $10_{H}$, $120_{H}$ and
$\ov{126}_{H}$, respectively the fundamental vector representation, the
three-index antisymmetric representation and the five-index antisymmetric and
anti-self-dual representation. $\ov{126}_{H}$ is necessarily complex,
supersymmetric or not; $10_{H}$ and $\ov{126}_{H}$ Yukawa matrices are
symmetric in generation space, while the $120_{H}$ one is antisymmetric.

Understanding fermion masses is easier in the Pati-Salam language of one of
the two maximal subgroups of SO(10), $G_{PS} =  SU(4)_{c}\times SU(2)_{L}
\times SU(2)_{R} $ (the other being $SU(5)\times U(1)$). Let us
decompose the relevant representations under $G_{PS}$

\begin{eqnarray}
{\bf 16} &=& (4,2,1) + (\bar 4,1,2) \\
{\bf 10} &=& (1,2,2) + (6,1,1) \\
{\bf 120}&=& (1,2,2) + (6,3,1) + (6,1,3) + (15,2,2) + (10,1,1) + (\ov{10},
1,1) \\
{\bf \ov{126}} &=& (\overline{10},3,1) + (10,1,3) + (15,2,2) + (6,1,1)
\end{eqnarray}
Clearly, the see-saw mechanism, whether type I or II, requires $\ov{126}$: it
contains both $(10,1,3)$ whose vev gives a mass to $\nu_{R}$ (type I), and
$(\ov{10},3,1)$, which contains a color singlet, $B-L= 2$ field $\Delta_{L}$,
that can give directly a small mass to $\nu_{L}$ (type II). A reader familiar
with the
SU(5) language sees this immediately from the decomposition under this group
\beq
{\bf \ov{126}} = 1 + 5 + 15 + \ov{45} + 50
\eeq 
The $1$ of SU(5) belongs to the  $(10,1,3)$ of $G_{PS}$ and gives a mass for
$\nu_{R}$, while $15$ corresponds to the $ (\overline{10},3,1)$ and gives the
direct mass to $\nu_{L}$. 

Of course, $ \ov{126}_{H}$ can be a fundamental field, or a composite of 
two $\ov{16_{H}}$ fields, or can even be induced as a two-loop effective
representation built out of a $10_{H}$ and two gauge $45$-dim representations.
In what follows I shall discuss carefully all three possibilities.

Normally the light Higgs is chosen to be the smallest one, $10_{H}$. Since
$\langle 10_{H}\rangle = \langle (1,2,2) \rangle_{PS}$ is a $SU(4)_{c}$
singlet, $m_{d} = m_{e}$ follows immediately, independently of the number of 
$10_{H}$ you wish to have.   Thus we must add either $120_{H}$ or  $
\ov{126}_{H}$
or both in order to correct the bad mass relations.  Both of these fields
contain $(15,2,2)_{PS}$, and its vev gives the relation $m_{e}= -3 m_{d}$.
 
As $ \ov{126}_{H}$ is needed anyway for the see-saw, it is natural to take
this first. The crucial point here is that in general $(1,2,2)$ and $(15,2,2)$
mix through $\langle (10,1,3)\rangle$\cite{Babu:1992ia} and thus the
 light Higgs is a mixture of
the two. In other words, $\langle (15,2,2) \rangle $ in $ \ov{126}_{H}$ is in
general non-vanishing \footnote{In supersymmetry this is not automatic, but
depends on the Higgs superfields needed to break SO(10) at $M_{GUT}$.}. It is
rather appealing that $10_{H}$ and $ \ov{126}_{H}$ may be sufficient for all
the fermion masses, with only two sets of symmetric Yukawa coupling matrices.

\subsection{An instructive failure}

Before proceeding, let me emphasize the crucial point of the necessity of
$120_{H}$ or $ \ov{126}_{H}$ in the charged fermion sector on  an instructive
failure: a simple and beautiful model by Witten\cite{Witten:1979nr}.
The model is non-supersymmetric and the SUSY lovers may place the blame for
the failure here. It uses $\langle 16_{H}\rangle$ in order to break $B-L$, and
the ``light'' Higgs is $10_{H}$.   Witten noticed an ingenious and
simple way of generating an effective mass for the right-handed
neutrino, through a two-loop effect which gives 
\beq
M_{\nu_{R}} \simeq y_{up} \left( \frac{\alpha}{\pi}\right)^{2} M_{GUT}
\eeq
where one takes all the large mass scales, together with $\langle 16_{H}
\rangle$, of the order $M_{GUT}$. Since $\langle 10_{H}
\rangle = \langle (1,2,2)_{PS} \rangle $ preserves quark-lepton symmetry, it
is easy to see that
\bea
M_{\nu} & \propto & M_{u} \nonumber \\
M_{e} & = & M_{d} \nonumber \\
M_{u} & \propto & M_{d} 
\eea
so that $V_{\rm lepton} = V_{\rm quark} = 1 $. The model fails badly. Is it yet
another example of beautiful theories killed by the  ugly facts of nature?

The original motivation of Witten was a desire to know the scale of
$M_{\nu_{R}}$ and increase $M_{\nu}$, at that time neutrino masses were
expected to be larger. But the real achievement of this simple, 
minimal SO(10) theory is the predictivity of the structure of $M_{\nu_{R}}$ and
thus $M_{\nu}$. It is an example of a good, albeit wrong theory: it fails
because it predicts.

What is the moral behind the failure? 
The main problem, in my opinion, was to ignore the fact that with only
$10_{H}$ already charged fermion masses fail. As I have argued repeatedly, one
needs a
 $ \ov{126}_{H}$ (or a $120_{H}$), one way or another, a lesson we keep in
what follows. 
    
 There is a nice way to save Witten's mechanism though. In order to work it leads
automatically to GUT scale heavy sfermions and light, order TeV, gauginos and
Higgsinos. I postpone its discussion to the section devoted to this, so called split
supersymmetry, scenario.

\section{Supersymmetric SO(10) GUT}

 In supersymmetry $10_{H}$ is necessarily complex and the
bidoublet $(1,2,2)$ in $10_{H}$ contains the two Higgs doublets of the MSSM,
with the vevs $v^{u}$ and $v^{d}$ in  general different: $\tan\beta\equiv
v^{u}/v^{d} \neq 1$ in general. In order to study the physics of SO(10), we
need to know what the theory is, i.e.~ its Higgs content.  There are two
orthogonal approaches to the issue, as we discuss now.

\subsection{\it Small representations}
The idea: take the smallest Higgs fields
(least number
of fields, not of representations) that can break SO(10) down to the MSSM and
give realistic fermion masses and mixings. The following fields are both
necessary and sufficient 
\beq
45_{H}, 16_{H }  + \ov{16}_{H}, 10_{H}\eeq
 It all looks simple and easy to deal with, but  
the superpotential   becomes extremely complicated. 
First, at the renormalizable level it is too
simple. The pure Higgs and the Yukawa superpotential at the
renormalizable level take the form
\bea
W_{H} &=& m_{45}  45_{H}^{2} + m_{16} 16_{H} \ov{16}_{H} +\lambda_{1} 16_{H}
\Gamma^{2} \ov{16}_{H} 45_{H} \nonumber \\ & &
m_{10} 10_{H}^{2} + \lambda_{2} {16}_{H} \Gamma {16}_{F} 10_{H} +
\lambda_{3}
\ov{16}_{H} \Gamma \ov{16}_{H} 10_{H}
\label{smallW}
\eea
\beq
W_{y} = y_{10} 16_{F} \Gamma 16_{F} 10_{H}
\label{smallWy}
\eeq
where $\Gamma$ stands for the Clifford algebra matrices of SO(10),
$\Gamma_{1}...\Gamma_{10}$, and the products of $\Gamma$'s are written in a
symbolic notation (both internal and Lorentz charge conjugation are omitted). 

Clearly, both $W_{H}$ and $W_{y}$ are insufficient. The fermion mass matrices
would be completely unrealistic and the vevs $\langle 45_{H} \rangle, \langle 
6_{H} \rangle, \langle \ov{16}_{H} \rangle $ would all point in the SU(5)
direction. Thus, one adds non-renormalizable operators 
\bea
\Delta W_{H} &=& \frac{1}{M_{Pl}} \left[  (45_{H}^{2})^{2}  + 
45_{H}^{4}
+ (16_{H} \ov{16}_{H})^{2} + (16_{H} \Gamma^{2} \ov{16}_{H})^{2}  + (16_{H}
\Gamma^{4} \ov{16}_{H})^{2}  \right .\nonumber \\ 
 & &
 + (16_{H}  \Gamma {16}_{H})^{2}  + (16_{H}
\Gamma^{5}  {16}_{H})^{2} + \{ 16_{H} \to \ov{16}_{H}\} \nonumber \\
& & \left .+ 16_{H} \Gamma^{4} \ov{16}_{H} 45_{H}^{2} + 16_{H} \Gamma^{3}
\ov{16}_{H} 45_{H} 10_{H} + \{ 16_{H} \to \ov{16}_{H}\} 
\right]
\eea
\bea
\Delta W_{y} &=& \frac{1}{M_{Pl}} \left[ 16_{F} \Gamma  {16}_{F} \, 16_{H} \Gamma
 {16}_{H}  +  \{ 16_{H} \to \ov{16}_{H}\} \right .  \nonumber \\
& & \left .  16_{F} \Gamma^{3}  {16}_{F}  45_{H} 10_{H} +  16_{F} \Gamma^{5} 
{16}_{F} \ov{16}_{H} \Gamma^{5}  \ov{16}_{H} \right ]
\eea
where I take for simplicity all the couplings to be unity; there are simply
too many of them.  The large number of Yukawa couplings means very little
predictivity.

The way out is to {\em add flavor symmetries} and to play the texture game and
thus
reduce the number of couplings. This in a sense goes beyond grand unification
and  appeals to new physics at $M_{Pl}$ and/or new symmetries (see e.g.~ 
\cite{Babu:1994kb}). 

To me, maybe the least appealing aspect of this approach is the loss of $R$
(matter) parity due to $16_{H}$ and $\overline{16}_{H}$; it must be postulated
by hand as much as in the MSSM. 

On the positive side, it is an asymptotically free theory and one can work in
the perturbative regime all the way up to $M_{Pl}$. While this
sounds nice, I am not sure what it means in practice. It would be crucial if
you were able to make high precision determination of $M_{GUT}$ or $m_{T}$,
the mass of colored triplets responsible for $d=5$ proton decay. The trouble
is that the lack of  knowledge of the superpotential couplings is sufficient
even in the minimal SU(5) theory to prevent this task; in SO(10) it gets even
worse.

Maybe more relevant is the fact that in this scenario $M_{R} \simeq
M_{GUT}^{2}/M_{Pl} \simeq 10^{13} -10^{14} \mbox{GeV}$, which fits nicely with the
neutrino masses via see-saw. Furthermore, see-saw can be considered ``clean'',
of the pure type I, since the type II effect is suppressed by $1/M_{Pl}$. Most
important, the $m_{b} \simeq m_{\tau}$ relation from (\ref{smallWy}) is
maintained due to small $1/M_{Pl}$ effects relevant only for the first two
generations.

 Now, the higher dimensional operators can be mimicked by the 
inclusion of singlets when they are integrating out (assuming them
much heavier that $M_{GUT}$, as expected by the gauge principle). This
paves the way for model building if one is willing to fine-tune their
masses to lie below  $M_{GUT}$, and this way one can get the double type I
see-saw formula (see e.g.\cite{Mohapatra:1986aw}).
  Recently Barr\cite{Barr:2003nn} has shown how in a particular
case
one can obtain the see-saw formula linear in $y_{D}$  (and not quadratic as
usual). Although obtained in a completely different manner,
this is what happens in the Witten's model, and thus in my opinion 
does not really represent a new type of see-saw.
Of course, this allows for different models of neutrino masses and
mixings. In order to stick to minimal theories, 
I refrain here from discussing this and similar proposals; this does not 
imply that they are without merit.

\subsection{\it Big is Better approach}
 The non-renormalizable operators
in reality mean invoking new physics beyond grand unification. This may be
necessary, but still, one should be more ambitious and try to use the
renormalizable theory only. This means large representations necessarily: at
least $\ov{126}_{H}$ is needed in order to give the mass to $\nu_{R}$ (in
supersymmetry, one must add $126_{H}$). The  consequence 
is the loss of asymptotic
freedom above  $M_{GUT}$, the coupling constants grow large at the scale
$\Lambda_{F}\simeq 10 M_{GUT}$\cite{Aulakh:2002ph} . To me this is a priori neither 
good nor bad,
but if it bothers you, you should skip the rest of the section. 

Once we accept large representations, we should minimize their number. The
minimal theory contains, on top of $10_{H}$, $126_{H}$ and $\ov{126}_{H}$,
also $210_{H}$\cite{mc,Clark:ai,Chang:1984uy,Aulakh:2003kg} with the
decomposition
\beq
210_{H} = (1,1,1)_{-} + (15,1,1)_{+} + (15,1,3) + (15, 3,1) + (6,2,2) +
(10,2,2) + (\ov{10},2,2)
\eeq
where the -(+) subscript denotes the properties of the color  singlets under
charge conjugation.

The Higgs superpotential is remarkably simple
\bea
W_{H} & = & m_{210} (210_{H})^{2}  +  m_{126}
\overline{126_{H}} 126_{H} + m_{10} (10_{H})^{2}  + \lambda (210_{H})^{3}  
\nonumber
\\
& &
 + \eta 126_{H }\overline{126}_{H} 210_{H} + \alpha 10_{H} 126_{H} 210_{H} 
 + \overline\alpha 10_{H}
\overline{126}_{H} 210_{H}
\label{WHso10}
\eea
and the Yukawa one even  simpler
\beq
W_{Y} =   y_{10}16_{F} \Gamma   16_{F} 10_{H} + y_{126}16_{F} \Gamma^{5}   16_{F}
\overline{126}_{H}
\label{Wyso10}
\eeq
Remarkably enough, this may be sufficient, without any higher dimensional
operators; however, the situation is not completely clear.

There is a small number of parameters: 3 + 6$\times$ 2 = 15 real Yukawa
couplings,
and 11 real parameters in the Higgs sector. In this sense the theory can be
considered as the minimal supersymmetric GUT in general\cite{Aulakh:2003kg}. As
usual,
I am not counting the parameters associated with the SUSY breaking terms.

The nicest feature of this program (and the best justification for the use of
large representations) is the following. Besides the $\langle (10,1,3) \rangle
$ which gives masses to the $\nu_{R}$'s,  also the  $\langle (15,2,2) \rangle 
$ in $\overline{126}_{H}$ gets a vev\cite{Clark:ai,Babu:1992ia}.
 Approximately
\beq
\langle 15,2,2 \rangle_{\ov{126}} \simeq \frac{M_{PS}}{M_{GUT}} 
\langle 1,2,2 \rangle
\eeq
with $M_{PS} =  \langle 15,2,2 \rangle $ being the scale of $SU(4)_{c}$
symmetry breaking. In SUSY, $M_{PS}\leq M_{GUT}$ and thus one can have correct
mass relations for the charged fermions.

What is lost, though, is the $b-\tau$ unification, i.e.~ with $\langle (15,2,2)
\rangle_{\ov{126}} \neq 0 $, $m_{b}= m_{\tau}$ at $M_{GUT}$ becomes an
accident. However, in the case of type II see-saw, there is a profound
connection between $b-\tau$ unification and a large atmospheric mixing angle.
 The fermionic mass matrices are obtained from (\ref{Wyso10})
\begin{eqnarray}
M_u &=& v_{10}y_{10}+v_{126}^u y_{126}\;, \nonumber \\
M_d &=& v_{10}y_{10}+v_{126}^d y_{126}\;, \nonumber\\
M_e &=& v_{10}y_{10}-3v_{126}^d y_{126}\;, \nonumber\\
M_{\nu_D}&=&v_{10}y_{10}-3v_{126}^u y_{126}\;, \label{fmasses}\\
M_{\nu_R}&=& y_{126} \langle ({10},1,3) \rangle\;,\label{rmass}\\
M_{\nu_L}&=& y_{126}  \langle (\ov{10},3,1) \rangle\;,\label{lmass}\\
\end{eqnarray}
where $\langle (\ov{10},3,1) \rangle \simeq M_{W}^{2}/M_{GUT} $ provides a
direct (type II) see-saw mass for light neutrinos. The form in (\ref{fmasses})
 is readily understandable, if you notice that $\langle (1,2,2) \rangle $ is a
$SU(4)_{c}$ singlet with $m_{q} = m_{\ell}$, and $\langle (15,2,2) \rangle $
is a $SU(4)_{c}$ adjoint, with $m_{\ell}=-3 m_{q}$ The vevs of the
bidoublets are denoted by $v^{u}$ and $v^{d}$ as usual. 

Now, suppose that type II dominates, or $
M_{\nu} \propto y_{126}  \propto M_{e} - M_{d}  $,
so that
\beq
M_{\nu} \propto M_{e} - M_{d}
\eeq
Let us now look at the 2nd and 3rd generations first. In the basis of diagonal
$M_{e}$, and for the small   mixing $\epsilon_{de}$
\beq
M_{\nu} \propto \left( \begin{array}{cc} m_{\mu} - m_{s} & \epsilon_{de} \\
\epsilon_{de} & m_{\tau} -m_{b}
\end{array}
\right)
\eeq
obviously, large atmospheric mixing can only be obtained for $m_{b} \simeq
m_{\tau}$\cite{Bajc:2002iw}.

Of course, there was no reason whatsoever to assume type II see-saw. Actually,
we should reverse the argument: the experimental fact of $m_{b} \simeq
m_{\tau}$ at $M_{GUT}$, and large $\theta_{\rm atm}$ seem to favor the type
type II see-saw. It can be shown, in the same approximation of 2-3
generations, that type I cannot dominate: it gives a small $\theta_{\rm atm}$
\cite{Bajc:2004fj}.
This gives hope to disentangle the nature of
the see-saw in this theory. As a check, it can be shown that the two types of
see-saw are really inequivalent\cite{Bajc:2004fj}.

 The three generation numerical studies supported a type II see-saw 
 \footnote{Type I can apparently be saved with CP phases, see
 \cite{japon}. For earlier work on type I see\cite{Lavoura:1993vz}}
with the
interesting prediction of a large $\theta_{13}$ and a hierarchical   neutrino 
mass spectrum\cite{Goh:2003sy}.  Somewhat better fits are obtained with a small
contribution of $120_{H}$ 
\cite{Bertolini:2004eq} or higher dimensional operators\cite{Dutta:2004wv}.

 I wish to stress an important feature of this programme. Since $126$
($\overline{126}$) is invariant under matter parity, R parity remains exact at
all energies and thus the lightest supersymmetric particle is stable and a
natural candidate for the dark matter.

\subsubsection{Mass scales}

In SO(10) we have in principle more than one scale above $M_{W}$ (and
$\Lambda_{SUSY}$): the GUT scale, the Pati-Salam scale where $SU(4)_{c}$ is
broken, the L-R scale where parity (charge conjugation) is broken, the scales
of the breaking of $SU(2)_{R}$ and $U(1)_{B-L}$. Of course, these may be one
and the same scale, as expected with low-energy supersymmetry. This solution
is certainly there, since the gauge couplings  of the MSSM unify
successfully and encourage the single step breaking of SO(10). 

Is there any room  for intermediate mass scales in SUSY SO(10)? It is
certainly appealing to have an intermediate see-saw mass scale $M_{R}$,
between $10^{12} - 10^{15} \mbox{GeV}$ or so. In the non-renormalizable case, with
$16_{H}$  and $\ov{16}_{H}$, this is precisely what happens: $M_{R}\simeq c
M_{GUT}^{2}/M_{Pl}\simeq c (10^{13} -10^{14}) \mbox{GeV}$. In the renormalizable
case, with $126_{H}$ and $\ov{126}_{H}$, one needs to perform a
renormalization group  study using unification constraints. While this is in
principle possible, in practice it is hard due to the large number of
fields. The stage has recently been set, for all the particle masses were
computed\cite{Bajc:2004xe,Fukuyama:2004xs}, and the preliminary
studies show that the situation may be under control\cite{Aulakh:2004hm}. It
is interesting that the existence of intermediate mass scales lowers the GUT
scale\cite{Bajc:2004xe,Goh:2004fy} (as was found before in models
with $54_{H}$
and $45_{H}$\cite{Aulakh:2000sn}), allowing for a possibly observable $d=6$ proton
decay. 

Notice that a complete study is basically impossible. In order to perform the
running, you need to know particle masses precisely. Now, suppose you stick to
the principle of minimal fine-tuning. As an example, you fine-tune the mass of
the $W$ and $Z$ in the SM, then you know that the Higgs mass and the fermion
masses are at the same scale
\beq
m_{H}= \frac{\sqrt{\lambda}}{g}m_{W} \, , \quad m_{f} = \frac{y_{f}}{g} m_{W}
\eeq
where $\lambda $ is a $\phi^{4}$ coupling, and $y_{f}$ an appropriate fermionic
Yukawa coupling. Of course, you know the fermion masses in the SM model, and
you know $m_{H} \simeq   m_{W}$. 

In an analogous manner, at some large scale $m_{G}$ a group $G $ is broken and
there are usually a number of states that lie at $m_{G}$, with masses
\beq
m_{i} = \alpha_{i} m_{G}
\eeq
where $\alpha_{i}$ is an approximate dimensionless coupling. Most
renormalization group studies typically argue that $\alpha_{i} \simeq O(1)$ is
natural, and rely on that heavily. In the SM, you could then take
$m_{H}\simeq m_{W}$, $m_{f}\simeq m_{W}$; while reasonable for the Higgs, it
is nonsense for the fermions (except for the top quark). 

In supersymmetry {\em all} the couplings are of Yukawa type, i.e.~
self-renormalizable, and thus taking $\alpha_{i}\simeq O(1)$ may be as wrong 
as taking all $y_{f}\simeq O(1)$. While a possibly reasonable approach when
trying to get a qualitative idea of a theory, it is clearly unacceptable when
a high-precision study of $M_{GUT}$ is called for.

\subsubsection{Proton decay}
As you know, $d=6$ proton decay gives $\tau_{p}(d=6) \propto M_{GUT}^{4}$,
while $(d=5)$ gives  $\tau_{p}(d=5) \propto M_{GUT}^{2}$. In view of the
discussion above, the high-precision determination of $\tau_{p}$ appears
almost impossible in SO(10) (and even in  SU(5)).  Preliminary studies
\cite{borutref} indicate fast $d=5$  decay 
as expected. 

You may wonder if  our renormalizable theory makes sense at all. After all, we
are ignoring
the higher dimensional operators of order $M_{GUT}/M_{Pl} \simeq 10^{-2} -
10^{-3}$. If they are present with the coefficients of order one, we can
forget almost everything we said about the predictions, especially in the
Yukawa sector. However, we actually know that the presence of $1/M_{Pl}$
operators is not automatic (at least not with the coefficients of order 1).
Operators of the type (in symbolic notation)
\beq
O_{5}^{p} = \frac{c}{M_{Pl}} 16_{F}^{4}
\eeq
are allowed by SO(10) and they give
\beq
O_{5}^{p} = \frac{c}{M_{Pl}} \left[( Q Q Q L ) + (Q^{c} Q^{c} Q^{c} L^{c}
)\label{30}
\right]
\eeq
These are the well-known $d=5$ proton decay operators, and for $c\simeq O(1)$
they give $\tau_{p}\simeq 10^{23} yr.$  Agreement with experiment requires
\beq
c \leq 10^{-6}
\eeq

Could this be a signal that $1/M_{Pl}$ operators are small in general?
Alternatively, you need to understand why just this one is to be so small. It is
appealing to assume that this may be generic; if so, neglecting $1/M_{Pl}$
contributions in the study of fermion masses and mixings is fully justified.

\subsubsection{Leptogenesis}

The see-saw mechanism provides a natural framework for baryogenesis through
leptogenesis, obtained by the out-of-equilibrium decay of heavy right-handed
neutrinos\cite{Fukugita:1986hr}. This works nicely for large $M_{R}$, in a sense too
nicely. Already type I see-saw works by itself, but the presence of the type II
term makes things more complicated \cite{hambyetalk}.
One cannot be a priori sure whether the
decay of right-handed neutrinos or the heavy Higgs triplets is responsible for
the asymmetry, although the hierarchy of Yukawa couplings points towards
$\nu_{R}$ decay. In the type II see-saw, the  most natural scenario is the
$\nu_{R}$ decay, but with the triplets running in the loops
\cite{Hambye:2003ka}. This and related issues are
now under investigation\cite{tpginprep}.
  
\section{Supersymmetry: is it really needed?}
 
  In the last two decades, and especially after its success
with gauge coupling unification, grand unification by an large got tied up
with low energy supersymmetry. This is certainly well motivated, since
supersymmetry is the only mechanism in field theory which controls the gauge
hierarchy. On the other hand, I hope to have convinced you that the right grand
unified theory should be based on SO(10), not SU(5). If so, gauge coupling
unification needs no supersymmetry whatsoever. It only says that there must be
intermediate scales\cite{intscales}, such as Pati-Salam $SU(4)_{c}\times
SU(2)_{L}\times SU(2)_{R}$ or Left-Right  $SU(3)_{c}\times SU(2)_{L}\times
SU(2)_{R}\times U(1)_{B-L}$ symmetry, between $M_{W}$ and $M_{GUT}$ (the
$SU(5)$ route is ruled out).  An oasis or two in the desert is always
welcome.

 Thus if we accept the fine-tuning, as we seem to be forced in the case
of the cosmological constant, we can as well study the ordinary, non-supersymmetric
version of the theory. In this context the idea of the cosmic 
attractors\cite{Dvali:2003br} 
as the solution to the gauge hierarchy becomes extremely appealing. It 
needs no supersymmetry whatsoever, and enhances the motivation for 
ordinary grand unified theories. In what follows I discuss some essential
features of a possible minimal such theory.

Let me stick to a purely renormalizable theory for the sake of simplicity
and predictivity.
The minimal such theory is based on  $210_{H}$, $\ov{126}_{H}$ (no need
for 
${126}_{H}$ as in SUSY) and a ``light'' $10_{H}$. In this case, the theory
is asymptotically free, and thus there is no advantage with small 
representations.
A purely renormalizable theory
can alternatively be built with $45_{H}$ and $54_{H}$ instead of $210_{H}$.
Notice that $45_{H}$ would not suffice: it turns out to preserve SU(5)\cite{Li:1973mq},
and
 $\ov{126}_{H}$ must preserve it in order not to break the SM symmetry. 

Intermediate mass scales help lower the masses of $\nu_{R}$, but create
potential problems for the charged fermions on the other hand. We have seen in
the supersymmetric version that the light Higgs is a mixture of $(1,2,2)$ in
$10_{H}$ and $(15,2,2)$ in  $\ov{126}_{H}$. This is crucial if
one is to get correct mass relations between down quarks and charged leptons.
These fields can mix through $(15,1,1)$ in $45_{H}$ or $210_{H}$, and
$(10,1,3)$ in   $\ov{126}_{H}$, either by the trilinear couplings
$(1,2,2), (15,2,2), (15,1,1)$
or the quartic ones
$
(1,2,2), (15,2,2), (15,1,1)^{2} $; $;   (1,2,2) , (15,2,2) ,
(10,1,3)^{2}$. 
In other words
\beq
\langle (15,2,2) \rangle \simeq \left(\frac{M_{I}}{M_{GUT}}\right)^{n}
\langle
(1,2,2)\rangle
\eeq
where $n=1,2$ for trilinear and quartic mixings respectively (which
depends on what the GUT scale fields are). Since  $\langle (15,2,2)
\rangle $ is needed for the second generation, with $m_{\mu}/M_{W} \simeq
10^{-3}$, we have the constraint
\beq
\left(\frac{M_{I}}{M_{GUT}}
\right)^{n}  \geq 10^{-3}
\eeq
This can be used to eliminate the single intermediate scale chain of 
symmetry breaking\cite{next}.

Another important difference with the SUSY situation lies in the Yukawa sector
where now, in the minimal theory, $10_{H}$ is real. This implies $v_{10}^{u} =
v_{10}^{d}$ and fitting fermionic masses and mixings becomes 
impossible\cite{next}. 

If it does fail, one could just add another $10_{H}$; this means unfortunately
another $y_{10}$. You can avoid it by postulating a Peccei-Quinn symmetry
\beq
16_{F} \to e^{i\alpha} 16_{F} \, , \quad 10_{H} \to e^{ -2 i\alpha} 10_{H} \, ,
\quad \ov{126}_{H} \to e^{ -2 i\alpha} \ov{126}_{H} 
\eeq
with $10_{H}$ now complex. This can give you naturally the axionic dark
matter,  at the expense of introducing additional $\ov{126}$ (or 126) in 
order to break both $B-L$ and $U(1)_{PQ}$. Although somewhat unappealing
and
against the rules of sticking
to the pure SO(10), the loss of neutralinos as the dark matter may necessitate
this. If you dislike this fact, simply work with two $10_{H}$'s and two
$y_{10}$'s.
Adding another $10_{H}$ is especially mild in the type II see-saw
since the relation  $ M_{d} - M_{e} \propto   M_{\nu}$ 
is independent of the number of $10_{H}$'s.
Thus $b-\tau$ unification is still connected with a
large $\theta_{\rm atm}$ as in the supersymmetric case.
  In recent years not much attention was devoted to
the ordinary SO(10), except for the work of the Napoli group (see e.g.~ 
Ref.\cite{Acampora:1994rh}). 

\section{Split Supersymmetry}

  If the need for the perturbative control of the weak scale is given up, there
appears an interesting alternative of split supersymmetry as I mentioned before.
This scenario which imagines a large supersymmetry breaking scale, with heavy
sfermions and light, order TeV, gauginos and Higgsinos has recently
attracted a lot of attention \cite{moresplit}.
Actually in the minimal SU(5) this is the only alternative to the low energy
supersymmetry; the only trouble is that SU(5) is not a good theory of massive
neutrinos. In SO(10) we have also an option of no supersymmetry as we have just
seen. How to know what the TeV effective physics, relevant for LHC, is? Is it
possible that a GUT answers the question?
    We recently came up with a theory that does just that\cite{Bajc:2004hr}. It is
based on a Witten's radiative seesaw scenario and its generalization to charged
fermions. All one needs is to have two $10_H$ fields, or one $10_H$ and one $120_H$
in the Yukawa sector. The main prediction of the theory is the split supersymmetry
with scalar masses of the order of the GUT scale (except of course for a light 
Higgs). To see that one needs to extend the radiative mechanism to a strongly
broken supersymmetric theory, with righthanded neutrino masses

\beq
M_{\nu_R}\approx\left({\alpha\over\pi}\right)^2Y_{10}  
{M_R^2\over M_{GUT}}f\left({\tilde{m}\over M_{GUT}}\right)\;,
\eeq

\noindent   
where $\tilde{m}$ is the scale of supersymmetry breaking, or in other
words the difference between the scalar and fermion masses of the  
same supermultiplet. This is valid only for $\tilde{m}$ not above
$M_{GUT}$. The function $f(x)\to 0$ when $x\to 0$ and
$f(x)={O}(1)$ if $x= O(1)$.

Due to the two loops suppression the only way to have
large enough righthanded neutrino masses is through single step
symmetry breaking $M_R\approx M_{GUT}$ and the large
$\tilde{m}\approx M_{GUT}$. The unification constraints with
no intermediate scale require then light gauginos and Higgsinos.
Thus, independently of the details
of the realistic Yukawa sector, one is forced to the split
supersymmetry picture.

\section{Summary and Outlook}

The see-saw mechanism has emerged in recent years as the simplest and the most
natural way of explaining small neutrino masses. It simply means adding
right-handed neutrinos  to the Standard Model, allowing them to
have large gauge invariant masses $M_{R}$ which break $B-L$ symmetry and through
Dirac Yukawa couplings $y_{D}$ with left-handed neutrinos give the latter
non-vanishing, but small masses. As appealing as this may be, as useless it is in
practice, The lack of knowledge of $M_{R}$ and $y_{D}$ leaves neutrino
masses arbitrary.  

 I have argued in this talk that  the most
natural framework for see-saw  is grand unification, a theory of large mass
scale
and the stage for the $q-\ell$ symmetry which can hopefully connect $y_{D}$ and
perhaps 
$M_R$
with quark Yukawa couplings.

If you accept this, then by now you should be convinced that the right GUT is
based on SO(10) gauge group. 
Basically all you ever wanted is there: 
right-handed neutrinos, Pati-Salam quark-lepton symmetry,
charge conjugation as a L-R symmetry and much more. The trouble is that
physics depends not only on the gauge symmetry, but almost as much on the
choice of the Higgs sector. It is here that the practitioners cannot
agree yet
and most attention is devoted to two rather orthogonal approaches. 
One insists
on perturbativity all the way to the Planck scale and chooses small
representations: $16_{H}$ (+ $\ov{16}_{H}$ in SUSY) and $45_{H}$. This program
then uses $1/M_{Pl}$ operators to generate the physically acceptable
superpotential; uses textures to simplify the theory and thus appeal to
physics beyond grand unification.

The other approach sticks to the pure SO(10) theory with no $1/M_{Pl}$
operators. This means large representation $\ov{126}_{H}$ (+$126_{H}$ in SUSY),
$210_{H}$; strong couplings in SUSY at $\lambda_{F} = 10 M_{GUT}$, but is
blessed with a small number of couplings and is a complete theory of matter
and non-gravitational  interactions. 
 This program is good
enough to be testable. Especially appealing is the version with the
type 2 seesaw since it offers a deep connection between a large atmospheric
mixing angle and b-tau unification; it furthermore predicts a large 1-3
leptonic mixing angle.
  
Strictly speaking in SO(10) one needs no supersymmetry at all, at least
not for the reasons of unification. If the un-naturalness of the small
Higgs mass is accepted, the nonsupersymmetric version with large
representation maintains all the good features discussed above, and remains
asymptotically free.

  Another, maybe even more intresting possibility, is a strongly broken
supersymmetry.
Namely, if one sticks to the minimal theory with a $16_{H}$ Higgs, then
a radiative seesaw mechanism works very nicely as long as supersymmetry
is split; one has a clear prediction of light gauginos and Higgsinos,
but superheavy, order GUT scale, sfermions. This is a rare example
when an inner structure of a  high-energy theory sheds light on the TeV physics
relevant for LHC without any assumptions about the naturalness.

I guess the main message of this short review is a caution when discussing the
see-saw mechanism. By itself, it is only an aesthetically appealing scenario
devoid of practical use. It makes sense to discuss it only in the context of a
well-defined theory based on firm physical principles. 
 
\subsection*{Acknowledgments}
  I am grateful to the organizers of this excellent meeting for the warm hospitality
and for the seesaw birthday party. I wish to acknowledge the great fun Rabi
Mohapatra and I always had doing physics, and especially while working on the
seesaw. I thank my friends Charan Aulakh, Borut Bajc, Pavel Fileviez-P\'erez, Thomas
Hambye, Alejandra Melfo and Francesco Vissani for enjoyable ongoing collaboration.
Thanks are also due to the members of the CMS group at FESB, Univ. of Split, Croatia
for their hospitality during the write-up of this talk. I am especially indebted to
Alejandra Melfo for her great help in the preparation of this review. This work is
supported by EEC (TMR contracts ERBFMRX-CT960090 and HPRN-CT-2000-00152).

\end{document}